# The role of new nuclear power in the UK's net-zero emissions energy system

James Price[a,*], Ilkka Keppo[b], Paul E Dodds[a]

[a] UCL Energy Institute, University College London, London, United Kingdom

[b] Department of Mechanical Engineering, Aalto University, Espoo, Finland

[*] Corresponding author: james.price@ucl.ac.uk

## Abstract

Swift and deep decarbonisation of electricity generation is central to enabling a timely transition to net-zero emission energy systems. While future power systems will likely be dominated by variable renewables (VRE), studies have identified a need for low-carbon dispatchable power such as nuclear. We use a cost-optimising power system model to examine the technoeconomic case for investment in new nuclear capacity in the UK's net-zero emissions energy system and consider four sensitivity dimensions: the capital cost of new nuclear, the availability of competing technologies, the expansion of interconnection and weather conditions. We conclude that new nuclear capacity is only cost-effective if ambitious cost and construction times are assumed, competing technologies are unavailable and interconnector expansion is not permitted. We find that BECCS and long-term storage could reduce electricity system costs by 5-21% and that synchronous condensers can provide cost-effective inertia in highly renewable systems with low amounts of synchronous generation. We show that a nearly 100% variable renewable system with very little fossil fuels, no new build nuclear and facilitated by long-term storage is the most cost-effective system design. This suggests that the current favourable UK Government policy towards nuclear is becoming increasingly difficult to justify.

## 1 Introduction

Many studies have highlighted the need for a concerted effort to rapidly transition power systems across the world from fossil fuels towards low-carbon technologies and particularly renewables (e.g. IPCC, 2018). This transition is considered a precursor to electrifying heat and transport and should be largely complete by 2035–2040 in a Paris-aligned world (IEA, 2021).

Wind and solar photovoltaic (PV) generation are already cost-competitive with fossil generation (IRENA, 2020) and their global capacities are growing rapidly; for example, the combined wind and solar shares were 29% and 24%[1] of total annual electricity generation in 2019 for Germany and the United Kingdom (UK), respectively. However, power production from VREs is driven by the weather and so it can vary rapidly in time and space. This leads to significant intermittency of supply and a marked paradigm shift from the dispatchable power systems of the recent past. Heuberger and Mac Dowell (2018) conclude that a 100% variable renewable UK system would run into significant operational difficulty, with periods of unmet demand and a lack of system inertia. This inertia is needed to stabilise system frequency and avoid instances of a large rate of change of frequency (RoCoF) which can lead to generator, load and interconnector disconnections.

---

[1] https://yearbook.enerdata.net/renewables/wind-solar-share-electricity-production.html

There is significant debate in the literature (Brown et al., 2018; Clack et al., 2017; Heard et al., 2017; Jacobson et al., 2017, 2015) regarding the future value of low-carbon dispatchable (LCD) power (nuclear, biomass, hydrogen and fossil fuels with carbon capture and storage) as the share of variable renewables increases. LCD plants can help to ensure that the supply of electricity is adequate (i.e. supply is able to meet demand during the normal operation of the system) and secure (i.e. to meet demand in light of unexpected contingency events such as a generator going offline). A number of studies have shown how LCD plants can reduce total system costs by limiting the need for the overcapacity of VREs and other flexibility options (Sepulveda et al., 2018; Zappa et al., 2019).

## 1.1 The value of new nuclear power generation

Of all LCD technologies, nuclear power generates perhaps the greatest controversy. Although nuclear plants have been operating in several countries for 60 years, the future of nuclear power across Europe is unclear, with France seeking to scale it back from around 70% of annual electricity generation today to 50% by 2035, and Germany intending to phase out nuclear altogether. Studies on the economics of new nuclear investments have reached differing conclusions, with little economic rationale in Sweden (Kan et al., 2020), but with nuclear a part of the lowest-cost future energy system in Switzerland (Pattupara and Kannan, 2016).

In the UK, which recently put in place a legally-binding commitment to achieve net-zero greenhouse gas emissions by 2050, the picture is just as unclear. The majority of the existing 9 GW of nuclear capacity is set to retire in the next few decades, only Sizewell B possibly still generating by 2050. Work has begun on a new nuclear power plant, Hinkley Point C, but costs and the timescale have escalated[2], and it is unclear whether other planned investments at Moorside (Cumbria), Wylfa (Anglesey) and Sizewell (Suffolk) will go ahead. The UK Government has agreed a guaranteed fixed price ( "strike price") for electricity produced from Hinkley Point C of £93/MWh[3], which is substantially higher than the £40/MWh strike price agreed for offshore wind farms opening in 2023-25[4]. Nevertheless, in the Energy White Paper published in December 2020, the UK Government has made clear its continued support for new nuclear capacity (BEIS, 2020a).

## 1.2 Aims and structure of this study

In this paper, we explore whether new nuclear beyond Hinkley Point C is likely to be economically-viable or necessary for an adequate and secure UK electricity supply in the context of a net-zero energy system. We focus on the third generation EPR pressurised water reactor design that is being built at Hinkley Point C and assume that any other third generation reactors would have similar costs and performance. We do not consider fourth generation reactors as these require substantial research and development and have not been constructed commercially, so have no credible cost or performance data.

Studies focusing on the future of the UK's electricity system in the context of its net-zero emissions objective either envisage a sizable build out (BEIS, 2020b; Daggash and Mac Dowell, 2019; Energy

---

[2] https://www.ft.com/content/fbc43de5-d3ae-49fd-9f5f-9e84f1db508d
[3] https://www.lowcarboncontracts.uk/cfds/hinkley-point-c
[4] https://www.gov.uk/government/publications/contracts-for-difference-cfd-allocation-round-3-results

Systems Catapult, 2020) or at least the replacement of existing nuclear capacity (CCC, 2020). However, these efforts all have some combination of the following methodological issues:

1. Important security of supply options are not considered. The ability for options other than LCD plants to provide inertia and frequency response services are neglected, ignoring the development of synchronous condensers and batteries.
2. Long-term storage with its potentially important role in providing a multitude of services is not considered.
3. The full role and value of interconnection with Europe is not considered.
4. They lack the necessary spatio-temporal resolution to appropriately model highly renewable electricity systems.

There is a relative paucity of work conducted within academia on this important question. We use a modelling framework that addresses the four points raised above and captures a range of sensitivities in terms of weather conditions, technological availability and costs.

This paper is structured as follows: in the next section we describe the model and methodology we use here, we follow that with a description and discussion of the results and finally we summarise the insights emerging from this study.

## 2 Methodology

### 2.1 highRES electricity system model

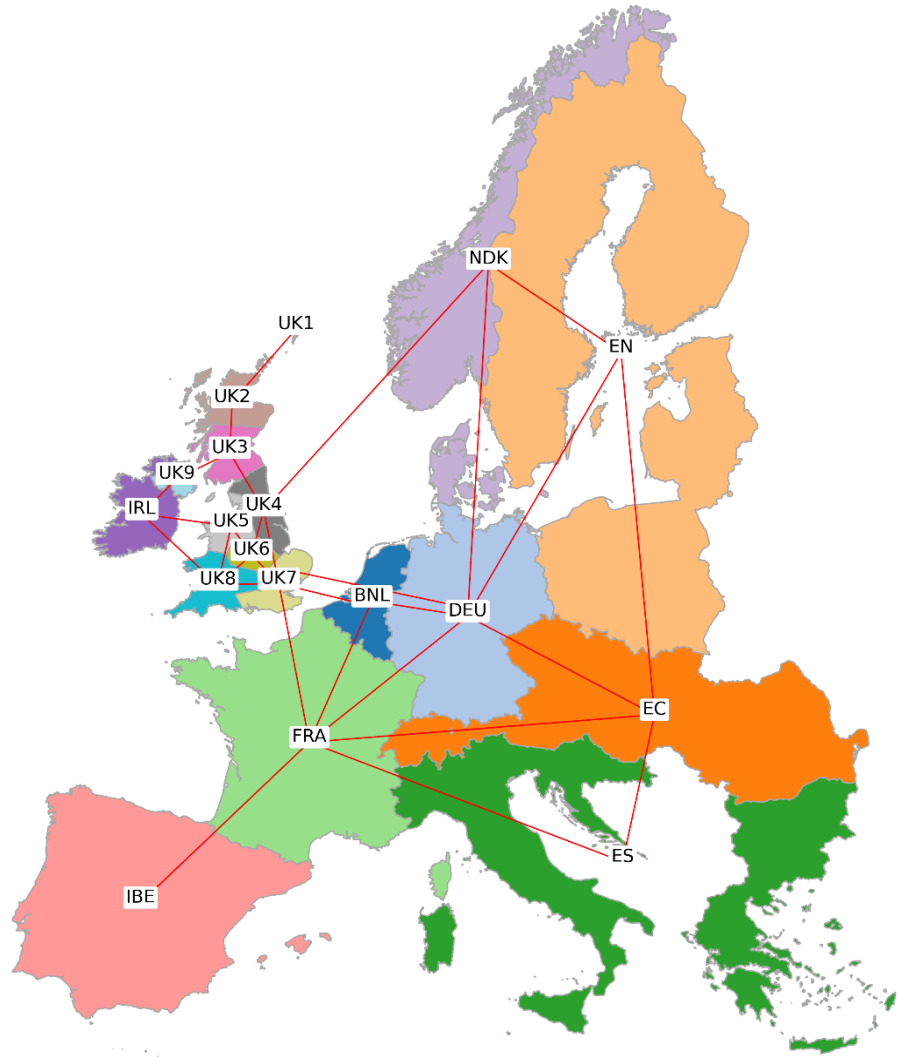

*Fig. 1 Map showing the spatial zones used in this work. The UK is composed of 9 interconnected zones with the remaining countries in Europe in various states of aggregation as described in the text.*

The implementation of the high spatial and temporal resolution electricity system model (highRES) we employ in this study is based on that used previously (see Price et al., 2020; Zeyringer et al., 2018) but spatially reconfigured and extended to cover the whole of the UK, broken down into 9 zones, and a further 27 European countries aggregated into an additional 9 zones as shown in Fig. 1. The rationale behind this aggregation is a trade-off between spatial detail and computational burden. Therefore, the UK, as the principle focus of this study, is modelled as multiple zones, with neighbouring countries mostly represented as individual nodes and those further away as regional groups. highRES is a cost-minimising model of this entire system written in the General Algebraic Modelling System (GAMS) language. It simultaneously optimises spatially-explicit capacity investment, based on annualised costs, and hourly dispatch in each of the 18 zones as well as interconnection between them. In addition, highRES also schedules hourly frequency response and operating reserve, which can both be provided by thermal generators and storage but not VREs (for details see Price et al., 2020).

We model capacity planning and dispatch for solar PV, on and offshore wind, nuclear, natural gas combined cycle turbines with CCS (NGCCGT-CCS) and open cycle turbines (NGOCGT). As part of our sensitivity analysis, we also make biomass with carbon capture and storage (BECCS) available to the model. All generation capacity online today is assumed to be retired by 2050 apart from Hinkley Point C and Sizewell B, resulting in a total of 4.5 GW of existing nuclear capacity. We note that while Sizewell B is currently due to retire in the mid-2030s, its owner EDF are targeting an extension of its operation lifetime until 2055.

We represent two energy storage options: grid-scale Lithium-Ion batteries with an 8-hour discharge duration, and in our sensitivity analysis we include long-term storage which uses hydrogen produced by electrolysis as the energy storage medium in salt caverns and $H_2$ burning CCGT and OCGTs for power output. The ratio between the energy and power components (i.e. the discharge duration) for the latter is free for the model to optimise. The technical energy storage potential for hydrogen storage in salt caverns for each country where available is taken from Caglayan et al. (2020).

For this work the model is free to expand transmission between zones within the UK while the capacity of interconnection between the UK and Ireland/Europe and between other European countries is a sensitivity dimension that we explore. Cost assumptions for all technologies are given in the Supplementary material.

For run-of-river, reservoir hydropower and pumped hydro storage we fix power capacities (i.e. no further investment is permitted) based on data from various sources including the ENTSO-E Transparency Platform[5] and Power Statistics[6], and national transmission system operators (TSO) in the first instance, with some gaps filled based on the JRC's hydropower database[7]. Energy storage capacities for reservoir and pumped storage are taken from Schlachtberger et al. (2017) and Geth et al. (2015) respectively. Inflow into run-of-river and reservoir plants is modelled following the methodology of Hörsch et al. (2018): run-off data from the European Centre for Medium Range Weather Forecasting ERA5 reanalysis used to set the hourly shape of the inflow time series across the year for each country, with total annual generation fixed to that reported by the Energy Information Administration (EIA)[8], in this case for 2012.

One important addition we make to the technologies represented in highRES for this study are synchronous condensers (SC) which provide inertia to the system while drawing a small amount of power from it. As discussed previously, in the real world inertia is needed to stabilise system frequency and limit RoCoF thereby contributing to system security. In our modelling we represent this via a constraint on hourly minimum inertia for the UK system, which is driven by an assumed RoCoF limit of 1 Hz/s which leads to a minimum inertia of 41.25 GWs given a largest loss of 1.65 GW (one unit of Hinkley Point C), and an hourly frequency response requirement that is linked to total system inertia (for details see Price et al., 2020). SC are well-established and used in Denmark, Italy and Australia because of the inertia they can provide to highly VRE powered systems with low amounts of synchronous generation. Indeed, the UK's TSO National Grid is currently in the process of procuring SC capacity to support renewable integration in the country's power system[9].

---

[5] https://transparency.entsoe.eu/
[6] https://www.entsoe.eu/data/power-stats/
[7] https://github.com/energy-modelling-toolkit/hydro-power-database
[8] https://www.eia.gov/international/data/world

The formulation of highRES has previously been described in detail in Price et al. (2020) and Zeyringer et al. (2018), so here we only describe additional equations developed for this study. These have been added to: i) model the unit operability of CCGTs and OCGTs coupled to long-term storage, and ii) to model the operation of reservoir hydropower. In the case of the former, we simply extend equations 1-9 from Price et al. (2020) to include long-term storage. For the latter, the formulation is as follows:

$$H_{h,z}^{level} = H_{h-1,z}^{level} + H_{h,z}^{inflow} - H_{h,z}^{gen} - H_{h,z}^{spill} \qquad (1)$$

$$H_{h,z}^{level} \leq H_{z}^{max\ level} \qquad (2)$$

$$H_{h,z}^{gen} \leq H_{z}^{cap} \times H^{avail\ fac} \qquad (3)$$

where $H^{level}$ is the amount of energy stored in each zones aggregated reservoir in GWh, $H^{inflow}$ is the natural inflow into each zones reservoir in GWh, $H^{gen}$ is the hourly power generation, $H^{spill}$ is the water spilled by the reservoir in each hour in GWh, $H^{cap}$ is the installed power generating capacity in each zone and $H^{avail\ fac}$ the availability factor of reservoir hydropower. The subscripts $h$ and $z$ are indices for hour and spatial zone, respectively.

Finally, we represent the planning and operation of thermal plants and long-term storage as clustered units following Palmintier (2014) whose integer decision variables are relaxed to be linear, which offers a significant computational speed up with only a limited loss of accuracy. For similar reasons, and given our focus on the UK, we opt to only apply these linearised unit commitment constraints, including the operating reserve and frequency response equations, to the UK system. The European system is then modelled using the standard linear approach as in Zeyringer et al. (2018), where technologies are represented as continuous lumps of capacity rather than units.

## 2.2   System boundaries

To model the design of UK power systems embedded within a wider energy system that achieves the countries net-zero emissions target by 2050, we constrain annual grid CO2 intensity to be ~2 gCO2/kWh based on the Balanced Pathway from CCC (2020). We also extend this definition of a net-zero compatible power system to cover Europe to ensure that the UK is not importing high-carbon electricity.

Hourly annual demand for 2050 for both the UK and the European zones is based on metered data from Open Power System Data[10] for 2012. As noted in Section 1, a substantial proportion of heat in buildings and road transport is likely to be electrified in the future, which would change the shape of the demand profile. Furthermore, the electrification of the former energy service demand, combined with the likely deployment of large amounts of VREs, will act to increase the coupling between supply and demand under a common meteorology. To capture both the change in total demand and the hourly shape, we take our 2012 demand profiles and develop a regression model for each country, based on an hourly extension of the approach used by Scapin et al. (2016) and informed by the methodology of Wang and Bielicki (2018), to separate out electricity demand into a temperature dependent and independent portions. We then augment this set of weather-independent profiles with an assessment of hourly electrified heat, ensuring that the weather driving both VRE supply and heat demand are the same, and electric vehicle demands. For further details, see the Supplementary material.

---

[9] https://www.nationalgrideso.com/news/latest-boost-stability-pathfinder-construction-flywheel-begins
[10] Open Power System Data. 2020. *Data Package Time series.* Version 2020-10-06. https://doi.org/10.25832/time_series/2020-10-06

For the UK, this methodology leads to annual heat in buildings demand that ranges from 74 to 107 TWh depending on the weather year, EV demand of 84 TWh, and total demand from 628 to 661 TWh. The total demand figure includes an additional 155 TWh from the Climate Change Committee's (CCC) Balanced Pathway which stems from manufacturing and construction, fuel supply, other demands and hydrogen production via electrolysis (not for use in the power sector). We assume these demands have a flat profile over the year.

When BECCS is available to the model we limit the biomass potential to ~61 TWh, as used by the CCC's Balanced Pathway in 2050, permitting ~17 TWh per year of electricity generation given a technology efficiency of ~28%. As shown by Fig 2.8 of CCC (2020), this is similar to today's use of biomass in the UK's power sector and ensures a degree of sustainability around this controversial technology.

## 2.3  Sensitivities

Here we consider a number of important sensitivity dimensions. First, given our focus on the prospects for nuclear power in the UK's net-zero aligned electricity system, we consider two capital cost levels for nuclear capacity. Both are based on analysis conducted for the Department for Business, Energy and Industrial Strategy (BEIS) by Leigh Fisher and Jacobs (2016) and consider a first of a kind (FOAK) overnight capital cost of 3927 £2010/kW and an nth of a kind (NOAK) cost of 3520 £2010/kW, or an approximate reduction of 10% for subsequent plants. This is to capture the uncertainty over whether the build out of new nuclear units will be relatively small in number, and hence a FOAK capital cost more appropriate, or more substantial, and so a NOAK costing more representative.

Second, we model the four technology availability scenarios listed in Table . We begin with a BASE case of technologies that are either available at scale today or, have been well demonstrated as for NGCCGT-CCS. We then add two more options that are at a somewhat earlier stage in their development: BECCS and long-term storage. BECCS is currently being trialled in the UK by Drax at their site in North Yorkshire and has been identified as being important for a net-zero energy system due to the negative emissions it can provide. Long-term storage is also being seen as potentially crucial, with BEIS recently announcing a competition with a total funding pot of £68 million to support the demonstration of long-term storage[11]. An "ALL" scenario then considers the simultaneous availability of both BECCS and long-term storage.

*Table 1 Definition of the technology availability scenarios used in this work*

| Scenario | Technologies available |
|---|---|
| BASE | Nuclear, NGCCGT-CCS, NGOCGT, Solar PV, On/offshore wind, Li-ion batteries, Synchronous Condensers |
| BECCS | BASE + BECCS |
| H2 | BASE + long-term storage (H2CCGT and H2OCGT for power production) |
| ALL | BASE + BECCS + long-term storage |

---

[11] https://www.gov.uk/government/publications/longer-duration-energy-storage-demonstration

Third, previous studies have highlighted the potentially critical role that the expansion of interconnection between countries can play in supporting the integration of high VRE shares. However, wider factors beyond simple least-cost solutions, such as Brexit, mean there are uncertainties as to the expansion of the UK's interconnection with Europe. We explore two situations in an attempt to capture this sensitivity. In the more conservative case, we fix the capacity of interconnection between countries to the 2027 net transfer capacities proposed by ENTSO-E's Mid-term Adequacy Forecast (MAF) in 2018. This means the UK has 16.5 GW of interconnection with Europe (including land-based transmission between Northern Ireland and Ireland) in our modelling (we update the MAF data for links between the UK to Norway and Denmark down from 4 GW to 2.8 GW which represents Viking Link and NSL). In addition, we limit the UK's maximum hourly net imports to 30% of the country's total hourly demand, noting that to date during 2021 this peaked at ~23% and will very likely rise in future as new interconnectors come online. The second, more ambitious case, allows highRES to optimise interconnection capacity up to a limit of 50 GW per link between countries. Here we allow hourly net imports into the UK to peak at 50% of total hourly demand. Note that in both situations the model can expand the transmission system within the UK to its optimal level.

Our fourth and final sensitivity dimension is the choice of weather year used to drive the production of wind and solar power and the demand for heat in our model. Past studies have demonstrated that different weather years lead to different optimal system designs (Zeyringer et al., 2018) and it may be the case that a relatively "poor" weather year, in which low wind/solar output coincides with cold temperatures and hence higher heating demands, would result in a more prominent role for nuclear power. However, running all of the 25 weather years (1993-2017) that highRES currently includes in conjunction with the other sensitivity dimensions discussed above would be computationally expensive. While there are many options for identifying the "worst", "average" and "best" years, here we take the route of running all the available weather years, one at a time, with the BASE technology options, FOAK nuclear costs and fixed interconnection capacity. From this we obtain a distribution of 25 total system levelised cost of electricity (system LCOE; that is total system cost divided by demand for the combined UK and European system). We assume the "worst" weather year (2010) for the modelled system has the highest LCOE, as it requires the highest combined expenditure for infrastructure planning (i.e. generators, storage, transmission and interconnection) and operational costs (fuel costs). The "average" (1995) and "best" (2014) years are then the median and lowest system LCOE, respectively.

Exploring all permutations of the sensitivity dimensions defined above results in 48 model runs (4 technology scenarios x 2 nuclear capex levels x 2 interconnection states x 3 weather years) which are executed using GAMS 27.2 and CPLEX 12.9 on a cluster computing environment.

## 2.4 UK system cost

As discussed by Kan et al. (2020), while the total cost of the combined system is a foundational output of optimisation models, here we wish to focus on the implications of our sensitivity dimensions for a specific country, the UK. To do that we draw from that work and make use of their definition of a total nodal system cost (equation 3 from that paper) and nodal LCOE (equation 4 from that paper). This approach captures all investments and operational costs incurred within the UK power system, considers the cost impact of the net electricity trade balance over the year and includes the share of congestion rent earned by the UK.

# 3 Results

## 3.1 Net-zero compatible system designs

First, we examine the cost-optimal role of nuclear and other LCD power plants for each of our 48 scenarios. Each scenario represents an adequate and secure electricity system for the weather year. Fig. 2 shows the installed capacity of all generator, storage and interconnection options by sensitivity case. We do not include wind and solar PV, which act to make the plots difficult to interpret owing to their total capacity of ~300 GW or more, and run-of-river and pumped hydro, whose capacity is fixed and does not change across the cases. The dashed line represents the total capacity of existing nuclear power expected to be online by 2050 (4.5 GW), thereby making any new investment clear. The results for the H2 scenario are not plotted as these are the same as the ALL technology set, as BECCS is not deployed in any cases where long-term energy storage is also available.

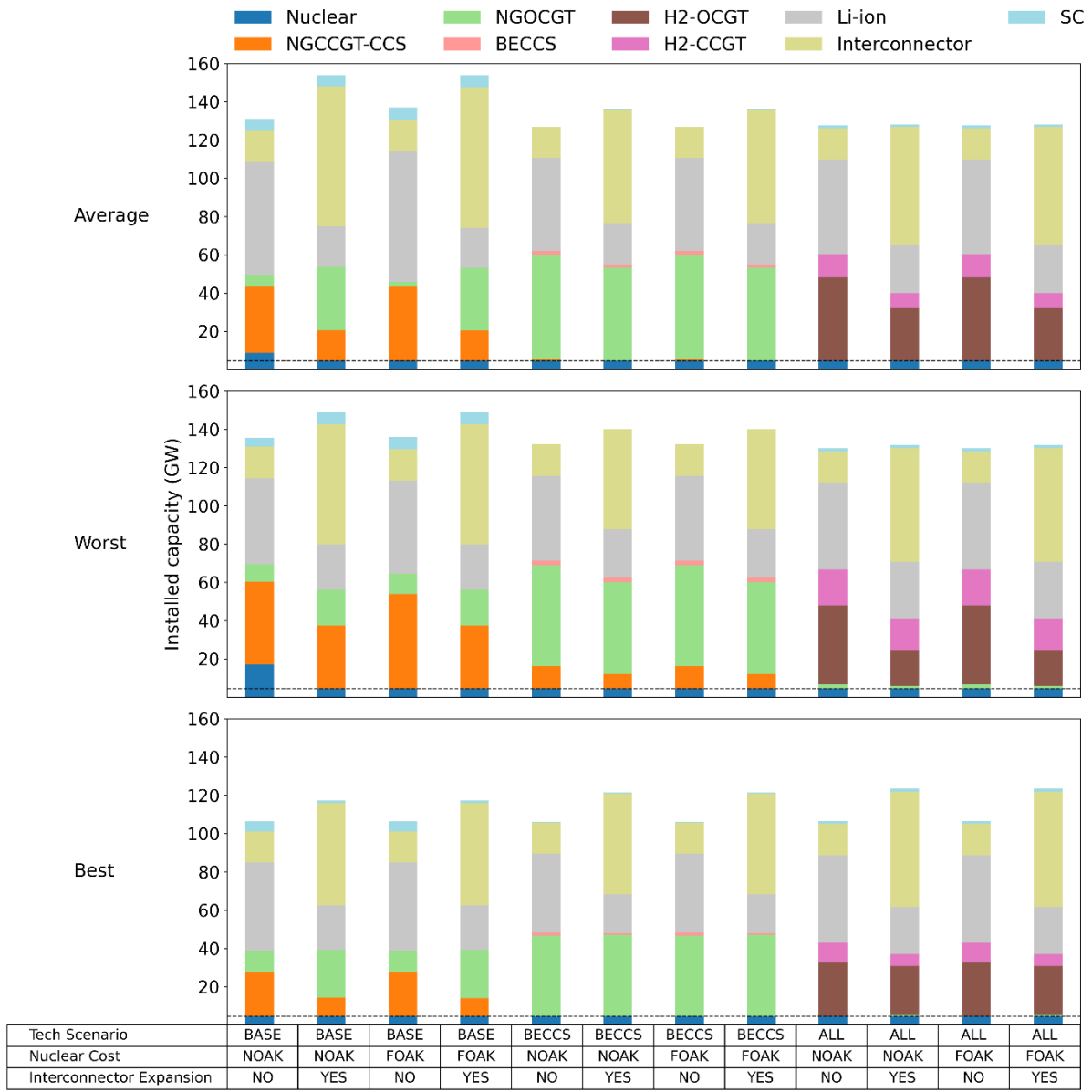

*Fig. 2 UK installed capacity by sensitivity with the three panels representing weather conditions. VREs are excluded to improve the readability of the figure due to their large installed capacities. Hydropower is also excluded as this does not change across the cases. The black dashed line shows the existing nuclear capacity in 2050 and allows the easy identification of new capacity being installed.*

In an average weather year, new nuclear is only cost-effective under the BASE technology scenario with NOAK nuclear capex and no interconnector expansion allowed, with 4 GW of new capacity installed. Permitting interconnector expansion under NOAK costs leads to no new nuclear capacity and the same is true under a switch to FOAK capital costs, regardless of the interconnector dimension. Furthermore, the remaining 8 cases plotted in the average weather year panel do not have any new build nuclear capacity. A similar pattern is observed for the worst weather year available in our sample of historic conditions but with a greater build out of nuclear than under average conditions, with the BASE-NOAK-NO case seeing an additional 13 GW of capacity. The best year sees no new build nuclear across all cases, which highlights the impact that the variability of weather within different weather years can have on VREs and, as a result, the wider power system design. Nevertheless, these panels clearly show that new nuclear power only features when the sensitivity dimensions are most in its favour, i.e. NOAK capex, no interconnection expansion and

BECCS and long-term storage being unavailable. If any of these change then new nuclear is not seen to be cost-effective.

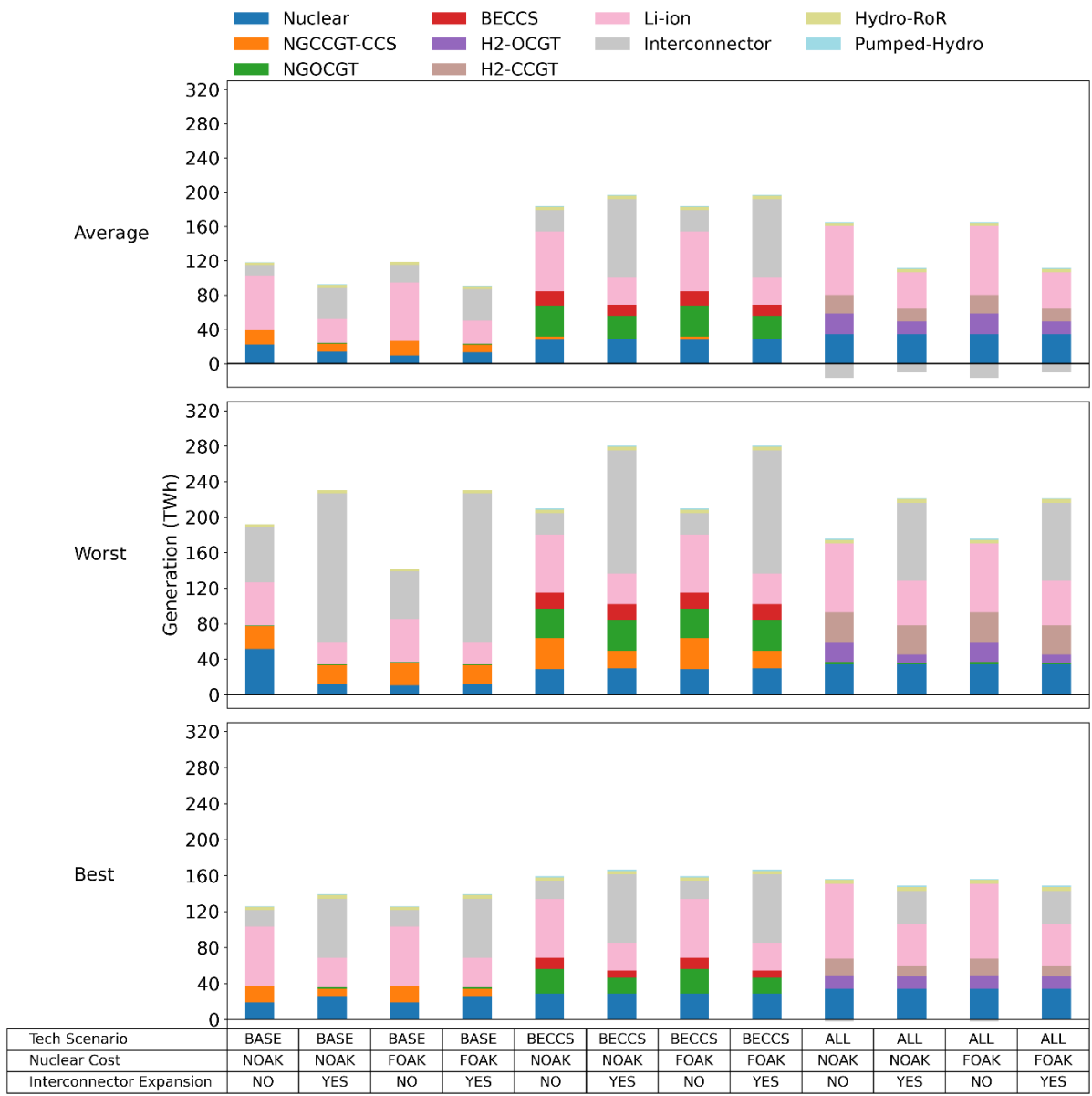

*Fig. 3 UK annual generation by sensitivity with the three panels representing the weather condition dimension. The country's net import position is also shown. Again, VREs are not shown for readability.*

Fig. 2 also shows a number of other insights. First, when interconnection expansion is permitted, the model consistently opts to substantially increase its capacity while reducing the role for batteries and, where available, long-term storage. This indicates the cost-effective nature of this form of flexibility for VRE integration. That said, this does lead to an import dependence for the UK under the BASE and BECCS scenarios of up to ~6% and ~14% of annual demand, respectively, in the case of average weather conditions. While cost-effective, it is a political decision as to whether this amount of annual net imports is acceptable from a domestic energy security standpoint.

Second, across all the weather conditions studied here, the addition of BECCS to the system drives a pronounced reduction in the deployment of NGCCGT-CCS in favour of NGOCGT and an increase in generation from the latter (see Fig. 3). This shows that when the annual $CO_2$ emissions budget can be extended by the negative emissions afforded by BECCS, open-cycle turbines represent a cheaper option to manage VRE intermittency and meet system adequacy/security requirements. Furthermore, our results indicate that at most 2.2 GW of BECCS is needed to drive this switch and to make new build nuclear cost ineffective.

Third, 1.5-6.3 GW of SC capacity is deployed in the BASE cases where new nuclear is not built and under every combination of the ALL technology scenario. This is because it provides inertia to support secure high VRE penetration and is seen to be more cost effective than nuclear power or BECCS to provide this service when long-term storage is available, despite consuming a small amount of electricity when online. This underscores the potentially important role this technology can have in future highly renewable systems which are inherently low on conventional synchronous generation.

Finally, when available, long-term storage displaces all or nearly all fossil generation in all weather conditions. The small amounts of NGOCGT deployed are in line with the assumption used here that a grid emissions intensity of ~2 $gCO_2$/kWh is compatible with a net-zero emissions energy system. Here H2-OCGT is generally seen to dominate over H2-CCGT in capacity terms, indicating that the system benefits from the greater flexibility and lower capital costs provided by the former, while still seeing a role for both. Furthermore, Fig. 3 shows that long-term storage also acts to reduce the import dependence, even to the extent of making the UK a small net exporter annually for average weather conditions. As this system design is chosen when all technology options are available, this also highlights that a close to fossil free, highly renewable system, enabled by long-term storage, is seen to be the most cost-effective scenario explored here.

## 3.2 UK system LCOE

Fig. 4 shows the UK system levelised cost of electricity (LCOE), which is the total UK nodal system cost divided by the total annual electricity demand, for all 4 of the sensitivity dimensions. We use a nodal system cost approach that accounts for capital, operational and trade costs associated with the node or collection of nodes of interest. The highest system costs are found for the BASE technology scenario and range from 52–70 £/MWh for the best to worst weather year, respectively. Permitting the model to expand interconnection leads to an average cost reduction of 6.9% for BASE cases.

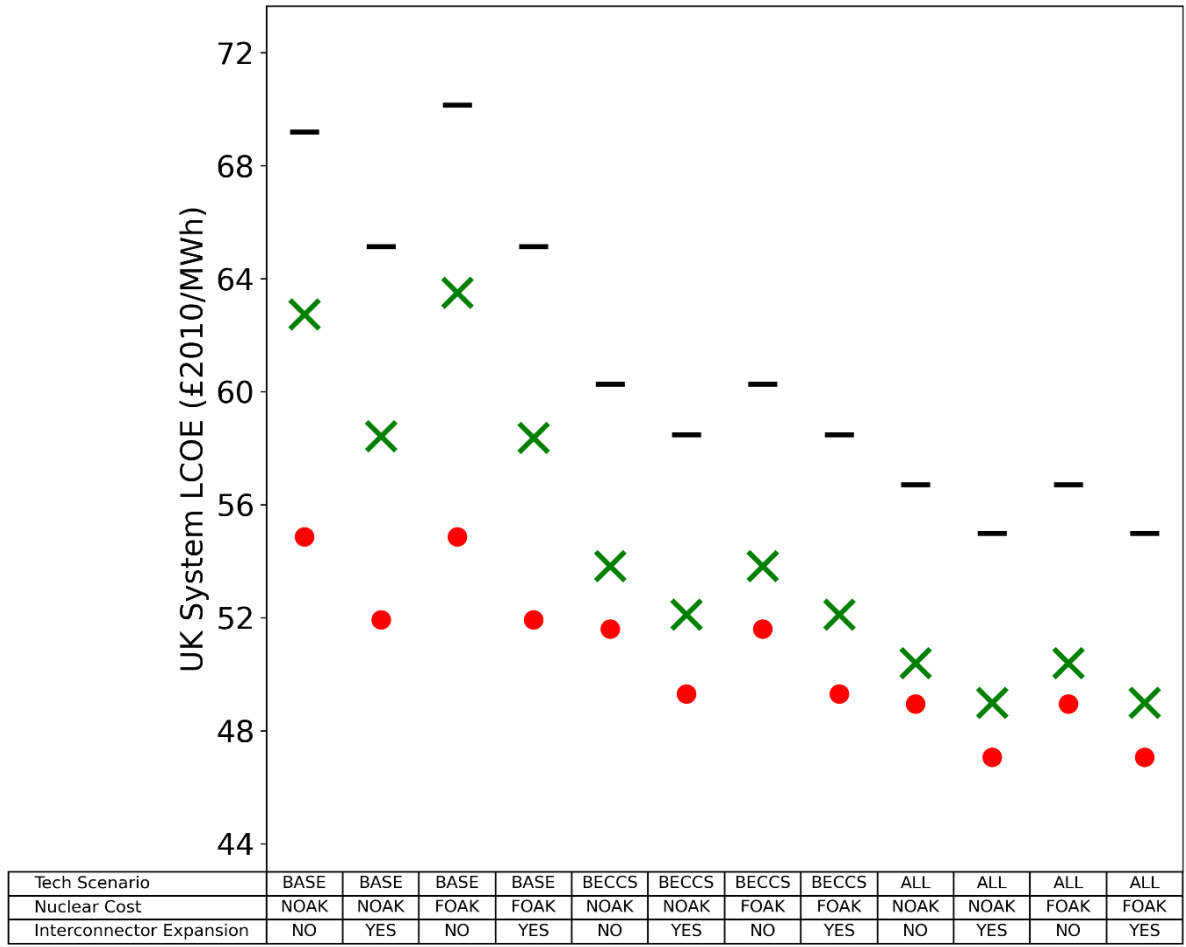

*Fig. 4 UK system LCOE by sensitivity. The weather dimension is shown using marker style and colour such as Best is red, Average is green and Worst is black.*

The introduction of BECCS leads to a 5–15% cost reduction across the sensitivity cases. The smallest reductions occur for the best weather conditions which are found to be essentially independent of nuclear cost or interconnector expansion. Reductions of 10–15% are seen for the average and worst weather conditions with the cases where interconnection expansion is not permitted consistently showing greater reductions compared to their equivalent scenario with expansion allowed. This implies the added flexibility coming from BECCS (i.e. extending the annual carbon budget resulting in a greater role for NGOCGT) provides relatively greater cost benefits, and is more valuable in poorer weather conditions and when interconnector capacity is fixed.

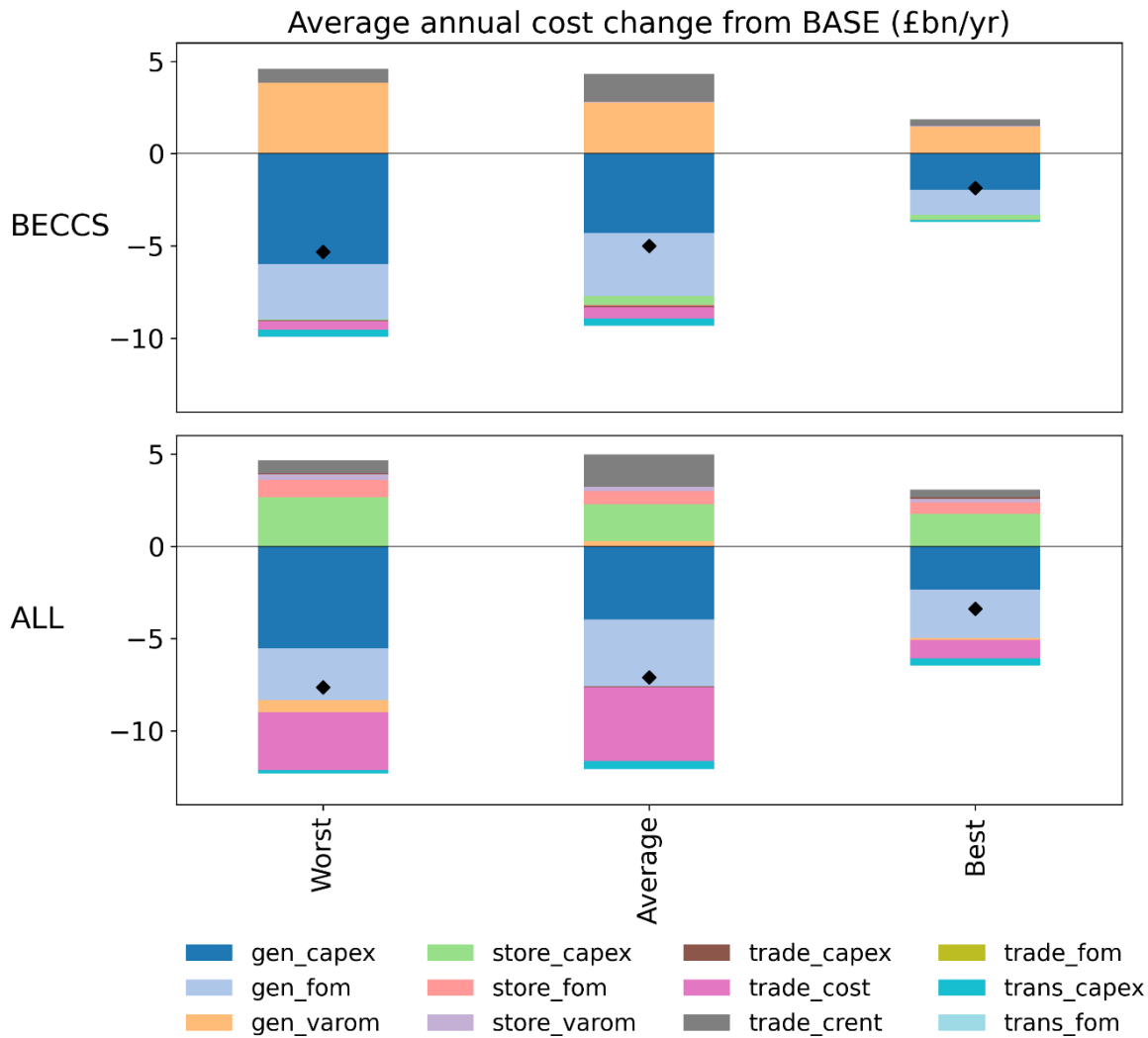

*Fig. 5 Average UK total system cost change from the BASE technology scenario by weather year. Average is across the nuclear and interconnection dimensions. Cost categories are "gen" for generation including SC, "store" for storage, "trade" for interconnection and "trans" for within country transmission. Cost terms are "capex" for capital cost, "fom" for fixed operating and maintenance cost, "varom" for variable operating and maintenance cost, "crent" for congestion rent and "trade_cost" the net of import costs – export revenues. Net change is denoted by black marker.*

The availability of long-term storage drives even greater cost reductions of 9–21%, again depending on the level of each sensitivity. Once more, the smallest reductions are seen for the best weather year with the average and worst conditions leading to large system LCOE decreases of 16–21%. More substantial reductions are found when further interconnection expansion is not allowed, again demonstrating that additional forms of system flexibility are most valuable when other options are unavailable.

The overall result of the system LCOE reductions identified above, that is greater cost reductions for the average and worst weather cases, is to compress the difference between the weather years. To understand what is driving this, in Fig. 5 we plot the average change in absolute costs when moving from the BASE to BECCS (upper panel) or ALL (lower panel) scenarios by weather year broken down into the cost components that make up the UK nodal system cost. Both panels show a reduction in annual generation capital and fixed O&M expenditure, which is larger in the average and worst weather years compared to the best year. The ALL technology scenario also shows a pronounced drop in costs associated with electricity trade, as the UK reduces its import dependence. Cost increases from the BASE scenario are largely related to the BECCS cases having greater generator variable O&M costs, driven by fuel costs for BECCS and natural gas, while the ALL cases see a rise in capex and fixed O&M costs associated with storage. Overall, as discussed above, both the BECCS and, even more so, ALL scenarios provide sizable annual cost savings compared to BASE.

### 3.3  UK VRE share in annual generation

Fig. 6 shows the share of VRE generation in total annual UK domestic generation by sensitivity and for each weather year. For the BASE technology scenario, VRE shares vary from 87% to over 95% depending on sensitivity, with the former being associated with, as might be expected, more pessimistic conditions for VREs, i.e. lower nuclear costs, worse weather and interconnection expansion not being permitted. Combining the results shown in Fig. 2 and Fig. 6 indicates that the spread in VRE penetration for the BASE-NOAK-NO case is likely driven by new nuclear capacity being deployed in the worst weather year.

For the BECCS cases, while there are differences between the different weather conditions, the share of generation from VREs is relatively consistent across the sensitivity options for a given weather year. Allowing the construction of new interconnector capacity leads to a small (~2%) drop in VRE share for the worst weather year while the average and best conditions modelled here see a small increase (~1%). This occurs because when interconnectors expand to leverage the geographic diversity of weather conditions across Europe the share of UK domestic VRE generation drops more than total UK domestic generation under the worst weather conditions. Nevertheless, these runs highlight how BECCS could support a UK power system that generates 80% or more of its electricity from VRE annually.

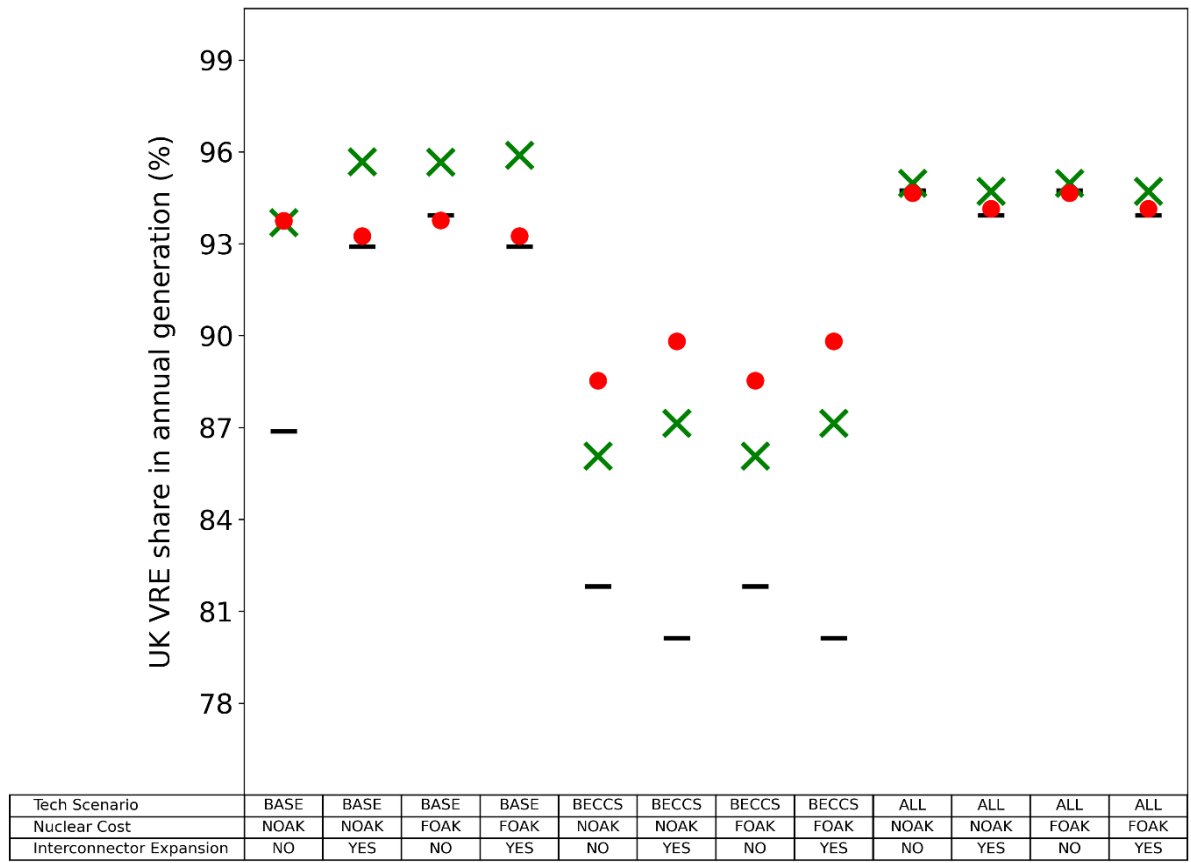

*Fig. 6 UK annual share of domestic generation from VRE by sensitivity. The weather dimension is again expressed using marker colour and style with Best as red circles, Average as green crosses and Worst as black dashes.*

The scenarios where long-term storage is available to the model see some of the highest VRE shares across the sensitivities at 94–95%, a finding that is essentially independent of annual weather conditions. This consistency shows that cost effective systems based on nearly 100% VRE generation, with no new build nuclear, can be robust to a variety of weather conditions when supported by long-term storage. Again, there is a small effect from further interconnector deployment resulting in a drop in VRE share, though at <1% this is very minor. These results highlight that long-term storage enables the integration of some of the highest amounts of domestic VRE generation in this study.

## 4   Discussion

Previous UK studies have envisaged a sizable build out (BEIS, 2020b; Daggash and Mac Dowell, 2019; Energy Systems Catapult, 2020) or at least the replacement of existing nuclear capacity (CCC, 2020). These strategies reflect comments in July 2021 from the UK Minister of State for Energy that "while renewables like wind and solar will become an integral part of where our electricity will come from by 2050, they will always require a stable low-carbon baseload from nuclear."[12] We have shown that new nuclear power generation is not necessary to provide electricity system adequacy and security. Even with challenging weather conditions, VRE generation could supply over 90% of total generation annually if coupled with technologies that provide or support system flexibility such as long-term storage, batteries, interconnector expansion and synchronous condensers. There are no substantial barriers to the use of these technologies in the UK.

---

[12] https://www.gov.uk/government/news/government-progresses-demonstration-of-next-generation-nuclear-reactor

We have not modelled demand-side measures in this study, which would offer another source of flexibility to the system and would help support the integration of VREs, due to significant uncertainty around the scale of their utilisation. We therefore consider our results a pessimistic case for VREs.

## 4.1 Cost assumptions

We assume an overnight capital cost for nuclear of 3927 $£_{2010}$/kW (FOAK) and 3520 $£_{2010}$/kW (NOAK). Assuming a 9% discount rate for the private sector (BEIS, 2020c), similar to Hinkley Point C (NAO, 2017), these are equivalent to LCOEs of 86 $£_{2010}$/MWh and 68 $£_{2010}$/MWh, respectively. These include interest during construction based on 10 and 6 year construction times for FOAK and NOAK respectively, where the latter should be regarded as very optimistic given construction lead times in the recent past (World Nuclear Association, 2021).

However, it is not clear that NOAK costs would be lower than FOAK costs. Negative learning-by-doing has been measured for the French and USA nuclear programmes (i.e. capital costs increased rather than reducing as the programme progressed). At best, costs have been reduced only very slightly in major programmes, for example in South Korea (Lovering et al., 2016). An assessment of US LCOEs concluded that cost estimates tend to be overly low and that historical analogues provide a better indication of likely costs (Koomey and Hultman, 2007); one reason is that nuclear reactors are the most prone to cost overruns as a percentage of budget and frequency (Sovacool, 2014). In contrast, solar PV and wind generation technology costs have reduced substantially over the last decade. Offshore wind strike prices for UK farms being built now are substantially lower (40 £/MWh) than the costs of new nuclear generation, even before accounting for nuclear cost overruns.

We have largely neglected socio-political considerations, for example public acceptance around nuclear, CCS and renewables. For example, the visual impact of onshore wind and solar PV can increase the cost of highly renewable systems (see e.g. Price et al., 2020). For nuclear, we have not included the cost of regulatory guarantees such as implicit public liability insurance in the case of an accident. We have also not included the cost of safely disposing spent nuclear fuel, which is also a political issue as the UK currently does not have a permanent safe storage facility.

The UK Government has proposed to fund new nuclear plants using a Regulated Asset Base (RAB)[13] model that would substantially reduce the discount rate and hence the strike price by making UK consumers liable for cost overruns and the risk of plants not being completed (Newberry et al., 2019). Given that this study has shown that a lower-cost low-carbon electricity system could be built using alternative technologies, it is questionable whether such favourable treatment can still be justified for nuclear generation, even before considering issues such as safely storing nuclear waste and hazards from low-probably but high-impact accidents.

## 4.2 Pathway to a net-zero energy system

Our approach to understanding the role of new nuclear in a net-zero emission energy system has been to design systems for a snapshot year with boundary conditions, in terms of emissions and demand, consistent with a net-zero energy system. We do not assess pathways to reach this system, in large part because of the necessary trade-offs between technical, temporal and spatial detail and time horizon needed to maintain computational tractability.

---

[13] https://www.gov.uk/government/consultations/regulated-asset-base-rab-model-for-nuclear

Renewable generation capacity has increased substantially over the last decade in Europe and there is confidence that the high penetration in our scenarios could be deployed. In contrast, new nuclear reactors have construction lead times of 6–12 years and a large specialised construction industry does not exist in the UK. Through not considering the pathway to a net-zero system, we have again most likely overestimated the potential role of nuclear generation.

### 4.3 Role of generation IV reactors

The UK Government has funded a £170 million Advanced Modular Reactor Demonstration Programme for Fourth Generation high temperature gas reactors (HTGRs)[14]. They have identified markets including hydrogen production and high temperature heat production to decarbonise heavy industry.

These reactors are much smaller than Third Generation reactors and are designed to be modular in nature in order to reduce capital costs through learning-by-doing. Existing nuclear reactor sizes have broadly increased over time to benefit from economies of scale (Grubler, 2010), as smaller reactors need similar investments in safety systems as large reactors. While HTGRs are designed with inherent safety features, it is not clear that they could be operated safely without a similar range of safety systems. Hence while individual reactors would be smaller and cheaper than Third Generation reactors, it is not clear that the LCOE would be lower. As such plants have not yet been developed, there is no credible cost data and we have not considered them in this study. There are, however, questions about both the cost and the deliverability of these reactors for a net-zero system.

## 5 Conclusions

We have sought to understand the role of new nuclear capacity in the UK's net-zero emissions energy system. Our sensitivity analysis on future UK power system designs has four key dimensions: nuclear capital costs; technology availability; interconnector expansion; and weather conditions. These dimensions were examined using a cost-optimising power system model of the UK and Europe with boundary conditions (electricity demand and $CO_2$ emissions) consistent with a net-zero energy system. The model was specifically configured to give high spatial and technical detail to the UK while still capturing the details of interconnection with low-carbon power systems across Europe. This analysis has generated a number of key insights:

- New nuclear capacity is found to be only cost-effective in the absence of BECCS, long-term storage and interconnector expansion and assuming NOAK nuclear capex with very ambitious construction times.
- BECCS reduces UK system LCOE by 5–15%, with greater savings seen for more challenging conditions (i.e. worse weather years; no interconnector expansion), as negative emissions facilitate the deployment of cheaper flexible assets.
- Long-term storage, modelled here as hydrogen generated from electrolysis and stored in underground salt caverns, can support 9–21% cheaper UK systems, with again more pessimistic assumptions for VREs leading to greater value from the flexibility it provides.

---
[14] https://www.gov.uk/government/news/government-progresses-demonstration-of-next-generation-nuclear-reactor

When both long-term storage and BECCS are available to the model, storage dominates and no BECCS is deployed.
- Synchronous condensers could have an important role in providing cost-effective inertia to support secure highly renewable systems in cases where synchronous generation is low. This includes BASE cases where no new nuclear is deployed and in all of the systems with long-term storage.
- The cost-optimal minimum share of annual generation from domestic VREs is found to be ~80% across all our scenarios with long-term storage consistently enabling ~94% share even in the worst weather year.

Each of these system designs account for the necessary amount of operating reserve, frequency response and minimum system inertia to support system adequacy and security. Taken together, these findings show that a nearly 100% variable renewable system with very little fossil fuels, no new nuclear and facilitated by long-term storage is the most cost-effective design presented here. Since a lower-cost, secure electricity system compatible with net zero could be built without new nuclear, the current favourable policy towards nuclear from the UK Government is becoming increasingly difficult to justify.

## Acknowledgements

This work was supported by the INNOPATHS project, which received funding from the European Union's Horizon 2020 research and innovation programme under grant agreement No 730403, and by the EPSRC SUPERGEN Energy Storage Hub project (EP/L019469/1).

# The role of new nuclear power in the UK's net-zero emissions energy system

## *Supplementary material*

**Cost assumptions**

*Table 1 Cost assumptions used in this work. Data is taken from the UK TIMES model (see [https://www.ucl.ac.uk/energy-models/models/uk-times](https://www.ucl.ac.uk/energy-models/models/uk-times)) unless otherwise stated via footnote. Overnight capital costs are combined with interest during construction costs before being annualised for input into highRES.*

| Technology | Overnight capex (£/kW) | Variable O & M (£/kWh) | Fixed O & M (£/kW) | Fuel costs (£/kWh) | Start-up cost (£/start) | Discount rate | Construction time (years) |
|---|---|---|---|---|---|---|---|
| Onshore wind | 745[1] | 0.005 | 20 | NA | NA | 5.2%[2] | 2 |
| Offshore wind | 1339[2] | 0.002 | 85 | NA | NA | 6.3%[2] | 3 |
| Solar | 243[1] | 0.001 | 5 | NA | NA | 5%[2] | 1 |
| NGCCGT-CCS | 1179 | 0.001 | 33 | 0.024 | 141,000[9] | 7.15% | 4 |
| Nuclear (FOAK) | 3927[3] | 0.004 | 63 | 0.006 | 195,000[9] | 9% | 10 |
| Nuclear (NOAK) | 3520[3] | 0.004 | 63 | 0.006 | 195,000 | 9% | 6 |
| NGOCGT | 309 | 0.001 | 14 | 0.024 | 3,100[9] | 7.15% | 2 |
| BECCS | 3844 | 0.004 | 104 | 0.014 | 141,000[9] | 7.15% | 4 |
| Li-ion batteries (Power) | 57[4] | 0 | 7 | NA | NA | 7.15% | 1 |
| Li-ion batteries (Energy) | 67[4] (£/kWh) | NA | 0 | NA | NA | 7.15% | 1 |
| Long-term storage (Power: Electrolyser + H2OCGT) | 439[5] | 0.005 | 14 | NA | 900[11] | 7.15% | 2 |
| Long-term storage (Power: Electrolyser + H2CCGT) | 691[5] | 0.004 | 30 | NA | 23,000[11] | 7.15% | 3 |
| Long-term storage (Energy: H2 Salt Cavern) | 0.027[6] (£/kWh) | | | | | 7.15% | 2 |
| HVAC line | 26[7] (£000/MW-100km) | | 0.5 (2% of capex) | | | 7.15% | 1 |

| Technology | | | | | |
|---|---|---|---|---|---|
| HVDC Subsea | 17.5[8] (£000/MW-100km) | | 0.35 (2% of capex) | | 7.15% | 1 |
| Synchronous condenser | 100[10] | | | | 7.15% | 1 |

[1] JRC (2018)
[2] BEIS (2020)
[3] Leigh Fisher and Jacobs (2016)
[4] Schmidt et al. (2019)
[5] H2OCGT and H2CCGT assumed to cost the same as natural gas fuelled OCGT and CCGT, electrolyser costs from IEA (2019).
[6] STORE&GO (2018)
[7] Parsons Brinckerhoff (2012)
[8] Zappa et al. (2019)
[9] Deane et al. (2015)
[10] Derived from AER (2019)
[11] Schill et al. (2017), only deprecation costs are assumed

**Technical assumptions**

*Table 2 Technical assumptions of technologies that are represented as units. H2OCGT and H2CCGT use the parameters from NGOCGT and NGCCGT-CCS respectively, emissions aside.*

| Technology | Inertia (s) | Unit size (MW) | Min stable generation (%) | Min up/down time (hours) | CO2 emissions (gCO2/kWh) |
|---|---|---|---|---|---|
| NGCCGT-CCS | 5[1] | 750 | 50[1] | 4/4[2] | 44 |
| Nuclear | 7[1] | 1650 | 50 | 24/8[2] | 0 |
| NGOCGT | 5[1] | 50 | 20[1] | 1/1[3] | 528 |
| BECCS | 5[1] | 500 | 50[1] | 6/6 | -1052 |
| Synchronous condenser | 6.5[4] | 200 | | | |

[1] Teng and Strbac (2017)
[2] Deane et al (2015)
[3] Heuberger et al (2018)
[4] ENTSO-E (2020)

**Demand modelling**

The functional form of the regression model used to subtract off existing electrified heating and cooling demands in each country is:

$$D_h = \sum_{i=1}^{12} \alpha_{h,i} Month_i + \sum_{j=1}^{7} \beta_{h,j} Day_j + \sum_k \gamma_{h,k} Holiday_k + \delta_h HDH_h + \varepsilon_h CDH + \epsilon_h$$

where $D_h$ is the hourly demand in each country which we assume is a sum of monthly dummy variables *Month* with *i=1* for January to *i=12* for December, day of week dummies *Day* with *j=1* for Monday to *j=7* for Sunday, holiday dummies *Holiday* with *k* varying depending on country, heating degree hours (HDH), cooling degree hours (CDH) and a residual term. The coefficients α, β, γ, account for differences by month, day of week and holidays respectively while δ and ε capture the assumed linear relationship between demand and HDH and CDH respectively. HDH and CDH are derived using ERA5 2m temperature data which is aggregated to country level using a population weighted average (via population data for 2011 from Eurostat[15]) and assuming cooling occurs from 20°C and heating from 15°C during the day and 10°C at night. These values were informed by the literature and by iteratively testing fit quality. Holiday dummy variables are country and year specific based on the Python module workalendar[16] with the regression model fit separately to each hour of the day. The 24 HDH and CDH coefficients are then averaged and used to subtract off hourly electricity demand from space heating and cooling resulting in a demand profile that is temperature independent. Here we assume that water heating does not correlate with outdoor temperature.

The next step is to add on electricity demand for space heating and cooling from the residential and commercial sectors in each country based on the weather year being modelled. For heating, we use data on useful energy demand (UED) for these sectors from the European Commission for 2012[17] which we adjust to the weather year being modelled by computing how the number of heating degree days (HDD) in 2012 compares to the number in that weather year at a country level. The annual UED is then further scaled to account for building fabric improvements in both sectors by 2050 based on Paardekooper et al. (2018) (assuming the average changes in UED for the 14 European countries they model are applicable to all the countries considered here). We then assume 90% of this space heating UED will be met by heat pumps and spread this annual figure out to the daily level on HDD derived using the ERA5 temperature data described previously. That is, days with a larger HDD are assumed to require a larger share of the annual UED. Daily UED is converted to electricity demand based on an estimate of the expected coefficient of performance for that day driven by the average daily temperature and using the relationship for air-source heat pumps from Staffell et al. (2012). Finally, hourly heating time series are obtained by using an average heat pump profile derived from Love et al. (2017) for the residential sector and Ruhnau et al. (2019) for the commercial sector to distribute out total daily demand.

For cooling, we take a simpler approach by only adding on the amount of annual electricity demand for cooling that was previously removed by the regression model. Nevertheless, the distribution of cooling demand throughout the year is still dependent on the weather year being modelled.

To model demand from electric vehicle charging we take the 600 TWh per year suggested by the European Commission's 1.5TECH scenario and share it out to the countries considered in that work assuming each country continues to use the same Europe wide share of final energy demand for road transport (for this we use 2011 data from ODYSSEE[18]). We then distribute out the annual demand for each country to each day of the year such that the same total demand is required each day but following different diurnal hourly profiles for week and weekend days taken from DESSTINEE demand model (Boßmann and Staffell, 2015).

In Fig. 1 we demonstrate our demand inputs to highRES by plotting a 10 day period around peak UK demand during the worst weather year (2010).

---

[15] https://ec.europa.eu/eurostat/web/gisco/geodata/reference-data/population-distribution-demography/geostat
[16] https://pypi.org/project/workalendar/
[17] https://ec.europa.eu/energy/studies/mapping-and-analyses-current-and-future-2020-2030-heatingcooling-fuel-deployment_en
[18] https://www.odyssee-mure.eu/

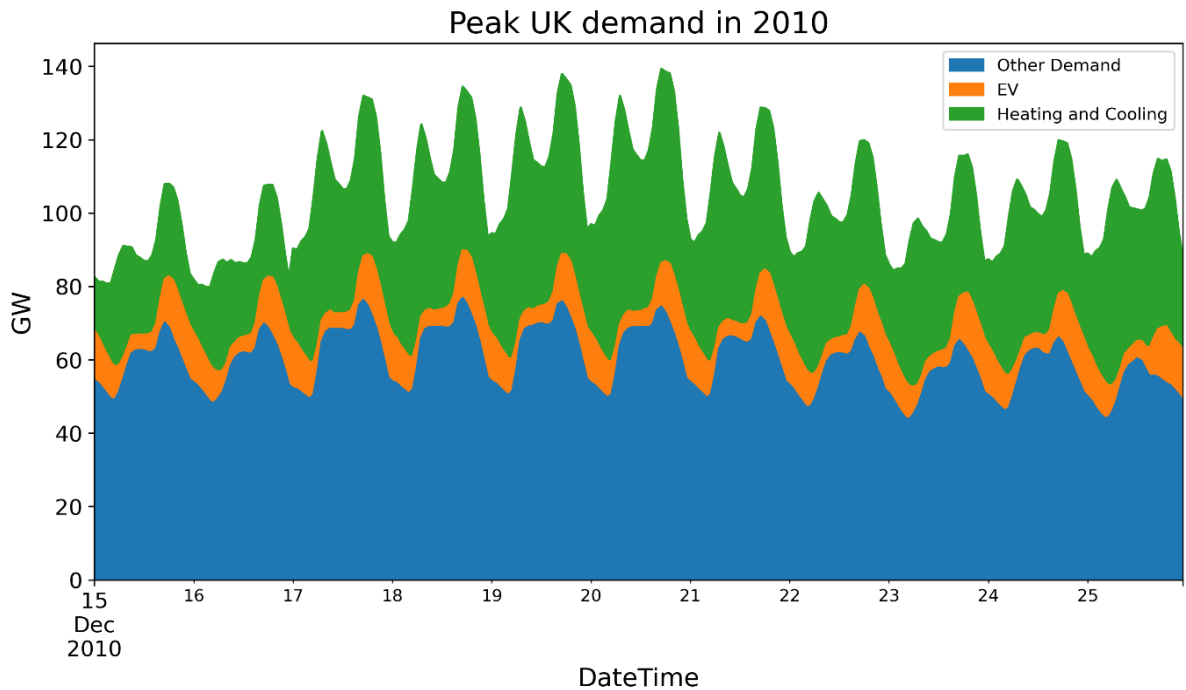

*Fig. 7 The 10 day period around the peak UK demand during the worst weather year (2010). Demand is split into that for heating and cooling, electric vehicles and everything else (other demand).*